\documentclass[11pt]{article}         

\usepackage{geometry}
\usepackage{graphicx}
\usepackage{amssymb,amsfonts,amsmath}
\usepackage{epstopdf}
\usepackage[footnotesize]{caption}
\usepackage{capt-of}

\usepackage{booktabs}
\usepackage{amsmath}
\usepackage{color}
\usepackage[normalem]{ulem} 

\begin{document}

\title{Poly(dA:dT)-rich DNAs are highly flexible in the context of DNA looping}

\author{%
Stephanie Johnson$^\dag$,\\Department of Biochemistry and
    Molecular Biophysics, \\California Institute of
    Technology, \\Pasadena, CA 91125\\
    Present address: Department of Biochemistry and Biophysics,\\ 
    University of California San Francisco, \\San Francisco, CA, USA\\
\and Yi-Ju Chen$^\dag$,\\Department of Physics, \\California Institute of
    Technology, \\Pasadena, CA 91125\\
\and Rob Phillips\footnote{To whom correspondence should be addressed.
Email: phillips@pboc.caltech.edu.
$^\dag$These authors contributed equally to this work.}, \\Departments of Applied Physics and Biology, \\California Institute of
    Technology,\\ Pasadena, CA 91125
}
\date{}

\maketitle

\begin{abstract}

Large-scale DNA deformation is ubiquitous in transcriptional regulation in prokaryotes and eukaryotes alike.  Though much is known about how transcription factors and constellations of binding sites dictate where and how gene regulation will occur, less  is known about the role played by the intervening DNA.  In this work we explore the effect of sequence flexibility on transcription factor-mediated DNA looping, by drawing on sequences identified in nucleosome formation and ligase-mediated cyclization assays as being especially favorable for or resistant to large deformations. We examine a poly(dA:dT)-rich, nucleosome-repelling sequence that is often thought to belong to a class of highly inflexible DNAs; two strong nucleosome positioning sequences that share a set of particular sequence features common to nucleosome-preferring DNAs; and a CG-rich sequence representative of high G+C-content genomic regions that correlate with high nucleosome occupancy \textit{in~vivo}.  To measure the flexibility of these sequences in the context of DNA looping, we combine the \textit{in~vitro} single-molecule tethered particle motion assay, a canonical looping protein,  and a statistical mechanical model that allows us to quantitatively relate the looping probability to the looping free energy.  We show that, in contrast to the case of nucleosome occupancy, G+C content does not positively correlate with looping probability, and that despite sharing sequence features that are thought to determine nucleosome affinity, the two strong nucleosome positioning sequences behave markedly dissimilarly in the context of looping.  Most surprisingly, the poly(dA:dT)-rich DNA that is often characterized as highly inflexible in fact exhibits one of the highest propensities for looping that we have measured.  These results argue for a need to revisit our understanding of the mechanical properties of DNA in a way that will provide a basis for understanding DNA deformation over the entire range of biologically relevant scenarios that are impacted by DNA deformability.

\end{abstract}

\newpage
\section{Introduction}

Although it has been known since the work of Jacob and Monod that genomes encode special regulatory sequences in the form of binding sites for proteins that modulate transcription, only recently has it become clear that genomes encode other regulatory features in their sequences as well.   Further, with the advent of modern
sequencing methods, it is of great interest to have a base-pair resolution understanding
of the significance of the entirety of genomes, not just specific coding regions
and putative regulatory sites.

One well-known example of other information present in
genomes is the different sequence preferences that confer nucleosome positioning \cite{Segal2009TG,Rando2010,Iyer2012}, with similar ideas at least partially
relevant in the context of architectural proteins in bacteria also \cite{Nishida2012}.
It has been shown both from analyses of sequences isolated from natural sources and from \textit{in vitro} nucleosome affinity studies with synthetic sequences that the DNA sequence can cause the 
relative affinity of nucleosomes for DNA to vary over several orders of magnitude, most likely due to the intrinsic flexibility, especially bendability, of the particular DNA sequence in question \cite{Widlund1999,Roychoudhury2000,Widom2001,Cloutier2004,Iyer2012}.
The claim that intrinsic DNA sequence flexibility determines nucleosome affinity has led not only to many theoretical and experimental studies on the relationship between sequence and flexibility \cite{Hogan1983,Peters2010,Olson2011,Olson2004,Morozov2009,Balasubramanian2009,Geggier2010}, but also to the elucidation of numerous sequence ``rules'' that can be used to predict the likelihood that a nucleosome will prefer certain sequences over others (summarized recently in \cite{Widom2001,Rando2010}). For example, AA/TT/AT/TA steps in phase with the helical repeat of the DNA, with GG/CC/CG/GC steps five base pairs out of phase with the AA/TT/AT/TA steps, are a common motif in both naturally occurring and synthetic nucleosome-preferring sequences \cite{Widom2001,Iyer2012}.  Similarly, the G+C content of a sequence and occurrence of poly(dA:dT) tracts have been very powerful parameters in predicting  nucleosome occupancy \textit{in vivo} \cite{Peckham2007,Tillo2009,Schwartz2009,Rando2010,Tillo2010}.  
Our aim here is to explore the extent to which these sequences, when taken beyond the context
of cyclization and nucleosome formation to another critical DNA deformation motif, exhibit similar effects on a distinct kind of deformation.

There has been an especially long history of the study of these intriguing sequence
motifs known as poly(dA:dT) tracts, in the context of nucleosome occupancy as well as many other biological contexts.   Such sequences, composed of 4 or more A bases in a row ($\mathrm{A}_n$ with $n\ge 4$) or two or more A bases followed by an equal number of T bases ($\mathrm{A}_n\mathrm{T}_n$ with $n\ge 2$), 
strongly disfavor nucleosome formation, both \textit{in vivo} \cite{Iyer1995,Segal2009,Field2008,Yuan2005} and \textit{in vitro} \cite{Rhodes1979,Kunkel1981,Prunell1982,Shrader1990,Anderson2001},  and are in fact thought to be one of the primary determinants of nucleosome positions \textit{in vivo} \cite{Rando2010,Segal2009}, with their presence upstream of promoters and in the downstream genes correlating with increased gene expression levels \cite{Struhl1985,Iyer1995,Segal2012NatureGenetics}.  
Poly(dA:dT) tracts show unique structural and dynamic properties in a variety of \textit{in vitro} and \textit{in vivo} assays (summarized recently in \cite{Haran2009,Segal2009}), with one of their hallmark characteristics being a marked intrinsic curvature \cite{Haran2009}.  There is evidence that poly(dA:dT) tracts may also be less flexible than other sequences \cite{Nelson1987,Suter2000,Vafabakhsh2012}, which is often given as the reason for their low affinity for nucleosomes, though there is some evidence that poly(dA:dT) tracts might actually be \textit{more} flexible than other sequences \cite{Hogan1983}.  It is clear, however, that some special property or properties of A-tracts leads them to be especially resistant to the deformations that are required for DNA wrapped in a nucleosome \cite{Segal2009,Haran2009}, and, indeed, to their important functions in several other biological contexts as well \cite{Haran2009}. 

In this work, we make use of sequences that, in the context of nucleosome formation and cyclization assays, appear to
be associated with distinct flexibilities as a starting point for examining the question of what sequence rules control deformations induced by a DNA-loop-forming transcription factor, as opposed to those induced in nucleosomes.  We have previously argued using two synthetic sequences that DNA looping does not necessarily follow the same sequence-dependent trends as do nucleosome formation and cyclization \cite{Johnson2012}.  Here we expand our repertoire of sequences to specifically test the generalizability of three sequence features known to be important in nucleosome biology and cyclization.  We  focus in particular  on the intriguing class of nucleosome-repelling, poly(dA:dT)-rich DNAs that are thought to be especially resistant to deformation, making use of a naturally occurring poly(dA:dT)-rich sequence that forms a nucleosome-free region at a yeast promoter \cite{Yuan2005}.  We note that the poly(dA:dT)-rich DNA we use here differs  from the \textit{phased} A-tracts that have been extensively characterized in the context of DNA looping, both \textit{in vivo} and \textit{in vitro} \cite{Crothers1990,Mehta1999,Edelman2003,Morgan2005,Haeusler2012,Cheema1999,Schulz2000,Serrano1991,Lilja2004}.  Phased A-tracts contain short poly(dA:dT) tracts spaced by non-A-tract DNAs such that the poly(dA:dT) tracts are in phase with the helical period of the DNA, generating globally curved structures that are known to significantly enhance DNA looping \cite{Crothers1990,Mehta1999,Edelman2003,Morgan2005,Haeusler2012}.  The poly(dA:dT)-rich sequence we examine here contains \textit{unphased} A-tracts that we do not anticipate to have a sustained, global curvature.

We compare the effects on looping of this poly(dA:dT)-rich DNA not only to the effects of two synthetic sequences we have previously studied, but also to those of two additional naturally occurring, genomic sequences: the well-known, strong nucleosome positioning sequence 5S from a sea urchin ribosomal subunit \cite{Simpson1983}, which, along with the 601TA sequence we previously studied, contains the repeating AA/TT/TA/AT and offset GG/CC/CG/GC steps that are common in nucleosome-preferring sequences; and one of the GC-rich sequences that are abundant in the exons and regulatory regions (\textit{e.g.}  promoters) of human genes, and that correlate with high nucleosome occupancy \textit{in vivo} \cite{Field2008, Schwartz2009,Tillo2010}.  The 5S sequence has been examined using both \textit{in vitro} cyclization and \textit{in vitro} nucleosome formation assays and, along with the two synthetic sequences E8 and 601TA \cite{Thastrom1999,Cloutier2004},  can be used as a standard for comparison between our and other \textit{in vitro} assays.  
The five sequences used in this work and their effects on nucleosomes are summarized in Table~\ref{tab:SeqDescript}.

\begin{table}[tbp]
\caption{
{\bf Naturally occurring and synthetic nucleosome-positioning or nucleosome-repelling sequences used in this study. }  }
    \label{tab:SeqDescript}
    \begin{center}
   \begin{footnotesize}
\begin{tabular}{ @{} l | l | l | l @{}} 
    \hline
    {\bf Sequence} & {\bf Species} & {\bf Genomic Position} 		& {\bf Nucleosome Affinity} \\
     {\bf Name}       & 		      &     &  \\
      		    &                  &                & \\
    \hline
    poly(dA:dT) & Budding yeast         & Chr III, 38745 -- 39785 bp  	& $\sim$3-fold \textit{in vivo} nucleosome \\
    	(``dA'')      &  (\textit{S.~cerevisiae}) &  	(Ref. \cite{Yuan2005})			& depletion relative to average  \\
	& & & genomic DNA (Fig. 2E of   	\\
	& & & Ref. \cite{Field2008}); $\sim 2~k_BT$ increase in \\
	& & & energy of  nucleosome formation  \\
	& & & \textit{in vitro} relative to 5S (Fig. 8D of  \\
	& & &Ref.~\cite{Field2008})  \textit{(estimates based on} \\
	& & & \textit{similar sequences)}\\
    \hline
    GC-rich & Human  & Chr Y, 4482107 -- 4481956 bp &  \textit{(not determined)} \\
    (``CG'')  & 	 &  (Ref. \cite{Field2008})   &  \\
     \hline
    5S & Sea~urchin		 &  20 bp-165 bp	& 1.6 $k_BT$ decrease in energy of   \\
   	& (\textit{L.~variegatus}) &  from the \textit{Mbo} II fragment   &  nucleosome formation compared \\
    	& 				 &  containing 5S rRNA gene & to E8 \textit{in vitro} (Ref.~\cite{Cloutier2004})\\
    	& 				 & 	(Ref. \cite{Simpson1983})		 &   \\
     \hline
        601TA & synthetic, strong  &  \textit{N/A}	& 3 $k_BT$ decrease in energy of   \\
   (``TA'')	& nucleosome   &  & nucleosome formation compared  \\
   & positioning sequence  & &to E8 \textit{in vitro} (Ref.~\cite{Cloutier2004})  \\
   & (Refs.~\cite{Lowary1998,Cloutier2004,Cloutier2005}) & & \\
     \hline
      E8 & synthetic random		 &  \textit{N/A}	& \textit{(used as a reference)}  \\
   	& (Refs.~\cite{Cloutier2004,Cloutier2005}) &   & \\
     \hline
    \end{tabular}
\begin{flushleft} The sequences described here were chosen because each has been found to have significant effects on \textit{in~vivo} nucleosome positions and/or \textit{in vitro} nucleosome affinities, as shown in the rightmost column.  The exception is the GC-rich sequence from humans: although its nucleosome affinity has not been directly determined either \textit{in vivo} or \textit{in vitro}, it is predicted to correlate with high nucleosome occupancy because of its high G+C content \cite{Tillo2009} and is occupied by a nucleosome(s) \textit{in vivo} according to micrococcal nuclease digestion \cite{Field2008}.   Two-letter abbreviations given in parentheses under each full sequence name will be used in figure legends in the rest of this work.  
\end{flushleft}
   \end{footnotesize}
      \end{center}
\end{table}

To measure the effect of these sequences on looping rather than nucleosome formation, we made use of a combination of an \textit{in vitro} single-molecule assay for DNA looping, called tethered particle motion (TPM)  \cite{Schafer1991,Yin1994,Finzi1995,nelson_tethered_2006}, with the canonical \textit{E. coli} Lac repressor to induce looping, and a statistical mechanical model for looping that allows us to extract a quantitative measure of DNA flexibility, called the looping J-factor, for the DNA in the loop \cite{Han2009,Johnson2012}.  We have recently demonstrated  \cite{Johnson2012} that this combined method offers a powerful and complementary approach to established assays that have been used to probe the mechanical properties of DNA, particularly at short length scales, to great effect, such as ligase-mediated DNA cyclization \cite{Shore1981,Crothers1992, Zhang2003, Cloutier2004, Du2005,Forties2009,Geggier2010,Geggier2011} 
and measured DNA end-to-end distance by fluorescence resonance energy transfer 
\cite{Kuznetsov2006,Vafabakhsh2012}. In particular, using the Lac repressor as a tool to probe the role of DNA deformability in loop formation allows us to examine the effect of sequence on the formation of shapes other than the roughly circular ones formed by cyclization and nucleosome formation, which we have argued may be an important caveat to discovering general flexibility rules from nucleosome formation and cyclization studies alone \cite{Johnson2012}. 

Interestingly, we find that the poly(dA:dT)-rich sequence that strongly excludes nucleosomes \textit{in vivo} \cite{Yuan2005} and that belongs to a class of sequences usually thought of as highly resistant to deformation is in fact the strongest looping sequence we have studied so far.  Moreover, the 5S and TA sequences, which share sequence features important to nucleosome formation (see Figures~\ref{fig:SIseqlist1} and~\ref{fig:SIseqlist2} in File S1 and Ref.~\cite{Widom2001}) as well as trends in apparent flexibility in \textit{in vitro} cyclization and nucleosome formation assays \cite{Lowary1998,Cloutier2004,Cloutier2005}, behave very differently from each other in the context of looping.  We also find that G+C content, a good predictor of nucleosome occupancy, is not likewise positively correlated with looping, and in fact our data suggest the G+C content and looping may be \textit{anti}correlated.  Taken together, these results strongly suggest that very different sequence rules determine DNA looping versus cyclization and nucleosome formation, possibly because of the protein-mediated boundary conditions that differ between looping geometries and nucleosomal geometries, and that the biophysical characteristics of poly(dA:dT)-rich DNAs and their biological functions may be more diverse and context-dependent than has been previously appreciated.

\section{Results}

\begin{figure}[tbp]
  \begin{center}
    \includegraphics[width=5in]{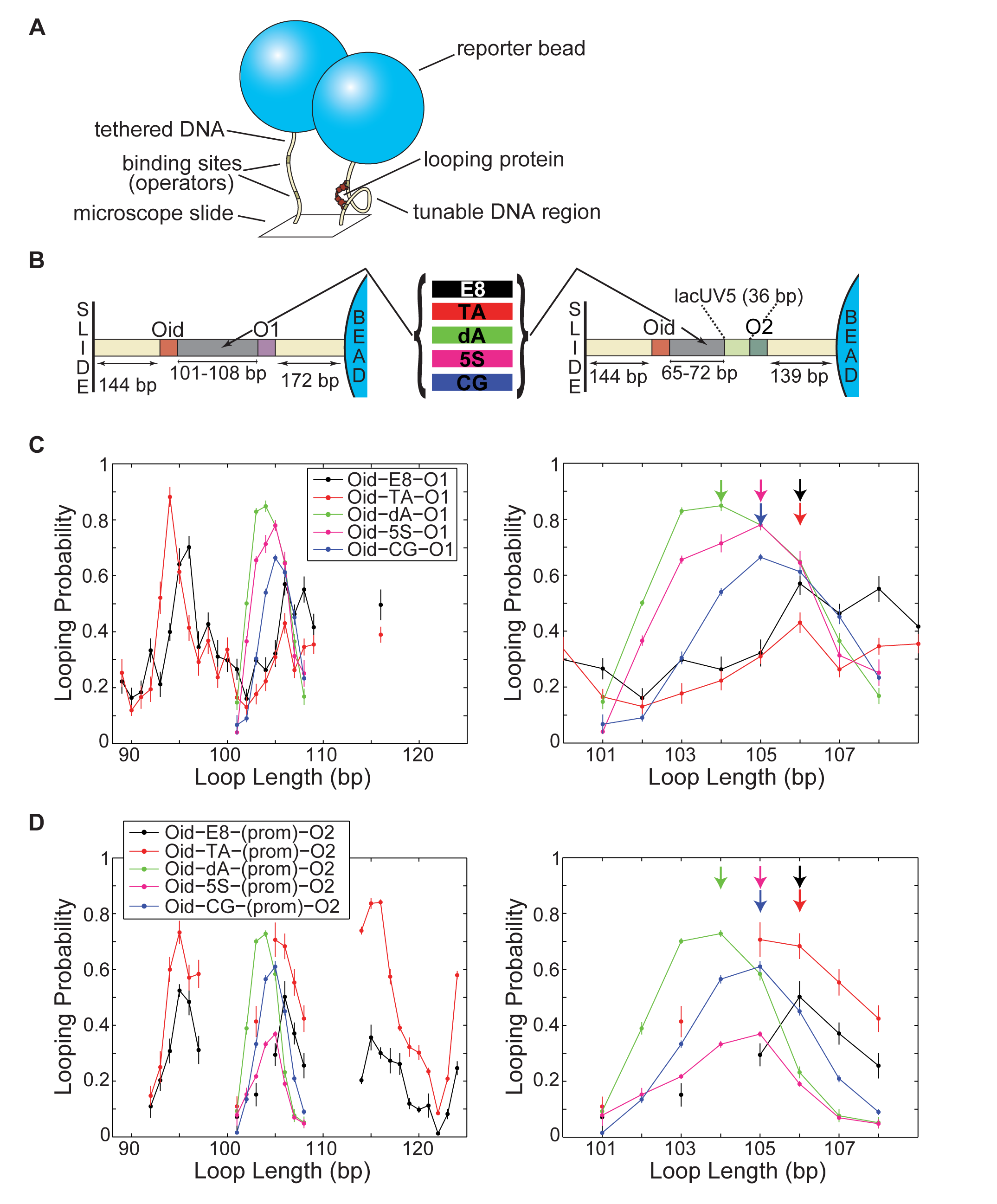}
    \end{center}
  \caption{
  {\bf Looping probability as a function of loop length and sequence.} 
 {\bf (A)} Schematic of the tethered particle motion (TPM) assay for measuring looping.  In TPM, a  bead is tethered to the surface of a microscope coverslip by a linear DNA.  The motion of the bead serves as a readout for the state of the tether: if the DNA tether contains two binding sites for a looping protein such as the Lac repressor, and the looping protein is present and binds both sites simultaneously, forming a loop, the motion of the bead is reduced in a detectable fashion \cite{Schafer1991,Yin1994,Finzi1995,nelson_tethered_2006}. The motion of the bead is observed over time, and the looping probability for a particular DNA is defined as the time spent in the looped (reduced motion) state, divided by the total observation time.
  {\bf (B)}  Schematic of the ``no promoter'' (left) and ``with-promoter'' (right) constructs used in this work. ``Loop length'' is defined as the inner edge-to-edge distance between operators (excluding the operators themselves, but including the promoter, if present). {\bf (C)}  Looping probabilities for the five sequences described in Table~\ref{tab:SeqDescript}, without the bacterial \textit{lac}UV5 promoter sequence as part of the loop. {\bf (D)} Looping probabilities for the same five sequences but with the promoter sequence in the loop.  Righthand panels in (C) and (D) show the same data as lefthand panels, except magnified around loop lengths 100-110 bp.  
  The five sequences do not all share the same maxima of looping (colored arrows),
  not even the TA and 5S sequences that share similar sequence features (see Figures~\ref{fig:SIseqlist1} and~\ref{fig:SIseqlist2} in File S1), though each sequence has the same maximum with and without the promoter (as far as can be determined with the current data; note that the with-promoter maximum for the TA sequence could be at 105 or 106, as those points are within error). 
  All E8 and TA data (in particular, those outside of the 101-108 bp range) were previously described in \cite{Johnson2012}. 
  }
  \label{fig:PloopTot}
\end{figure}

Our experimental approach to examining the effect of DNA sequence on looping combines an \textit{in vitro} single-molecule assay for DNA looping, called tethered particle motion (TPM) \cite{Schafer1991,Yin1994,Finzi1995,nelson_tethered_2006}, with a statistical mechanical model that allows us to extract biological parameters from the single-molecule data \cite{Han2009,Johnson2012}.  As shown schematically in Fig.~\ref{fig:PloopTot}(A), in TPM, a microscopic bead is tethered to a microscope coverslip by a linear piece of DNA, with the motion of the bead serving as a reporter of the state of the DNA tether:  the formation of a protein-mediated DNA loop in the tether reduces the motion of the bead in a detectable fashion \cite{Schafer1991,Yin1994,Finzi1995,nelson_tethered_2006}.  We use the canonical Lac repressor from \textit{E. coli} to induce DNA loops.  Because more readily deformable sequences allow loops to form more easily, we can quantify sequence-dependent DNA flexibility by quantifying the looping probability, which we calculate as the time spent in the looped state divided by the total observation time (see Methods for  details).  

More precisely, our statistical  mechanical model (described in the Methods section)  allows us to extract a parameter called the looping J-factor from looping probabilities \cite{Johnson2012}.  The J-factor is the effective concentration of one end of the loop in the vicinity of the other, analogous to the J-factor measured in ligase-mediated DNA cyclization assays \cite{Shore1981,Jacobson1950}, and is mathematically related to the energy required to deform the DNA into a loop, $\Delta F_\mathrm{loop}$, according to the relationship:
\begin{equation}
J_{\mathrm{loop}} = 1~\mathrm{M}~e^{-\beta\Delta F_{\mathrm{loop}}},
\end{equation}
where $\beta=1/(k_BT)$ ($k_B$ being Boltzmann's constant and $T$ the temperature).  A higher J-factor therefore corresponds to a lower free energy of loop formation.  In the case of cyclization, where the boundary conditions of the ligated circular DNA are well understood, the J-factor can be expressed in terms of parameters describing the twisting and bending flexibility of the DNA, and its helical period \cite{Shimada1984,Shore1983,Peters2010,Geggier2010}.  However, in the case of DNA looping by the Lac repressor, where the boundary conditions are not well known (summarized in Fig. 4 of \cite{Johnson2012}), an expression for the looping J-factor in terms of the twist and bend flexibility parameters of the loop DNA has not been described.  Nevertheless, by measuring the J-factors for different sequences, we can comparatively assess the effect of sequence on the energy required to deform the DNA into a loop, and thereby gain insight into the sequence rules that control this deformation.

\subsection{Loop sequence affects both the looping magnitude and the position of the looping maximum.}

Given that 5S and TA share both sequence features and similar trends in apparent flexibility in the contexts of nucleosome formation and cyclization \cite{Lowary1998,Cloutier2004,Cloutier2005}, we expected these two sequences to behave similarly to each other in the context of looping.  On the other hand, since poly(dA:dT)-rich sequences are supposed to assume such unique structures as to strongly disfavor nucleosome formation \cite{Segal2009,Haran2009}, while high GC content is one of the strongest predictors of high nucleosome occupancy \cite{Tillo2009,Field2008}, we expected these two sequences to behave very differently from each other in the context of looping. Given the common assumption that poly(dA:dT)-rich DNAs are highly resistant to deformation, we especially did not expect to observe much, if any, loop formation with the poly(dA:dT)-rich, nucleosome-repelling sequence.

As shown in Fig.~\ref{fig:PloopTot}, none of these expectations were  borne out. TA and 5S do not behave similarly, nor do CG and poly(dA:dT) behave especially dissimilarly, nor does poly(dA:dT) resist loop formation. 
Moreover, the behavior of these special nucleosome-preferring or nucleosome-repelling sequences is dependent on the larger DNA context, in that the addition of the 36-bp bacterial \textit{lac}UV5 promoter sequence to these roughly 100-bp loops changes the relative looping probabilities of the five sequences (see Methods for the rationale behind the inclusion of this promoter).  Without this promoter sequence (Fig.~\ref{fig:PloopTot}(C)), the two synthetic sequences, E8 and TA, exhibit comparable amounts of looping, while the three natural sequences, including both 5S and poly(dA:dT), all loop more than either E8 or TA.  With the promoter (Fig~\ref{fig:PloopTot}(D)), however, TA loops more than E8, but 5S \textit{less} than either E8 or TA.  Both with and without the promoter the supposedly very different GC-rich and poly(dA:dT)-rich DNAs loop more than the random E8 sequence.  The looping probabilities of the poly(dA:dT) sequence are especially surprising---instead of looping very little, as we expected, this sequence loops more than any other sequence without the promoter and a comparable amount to TA with the promoter.

These five sequences differ not only in looping probability, but also in the loop length at which that looping is maximal: the poly(dA:dT) sequence is maximized at 104~bp, the 5S and CG sequences at 105~bp, and the E8 and TA sequences at 106~bp.  These different maxima could be explained by different helical periods for these five DNAs, though without more periods of data we cannot definitively quantify their helical periods.  In the case of the poly(dA:dT) sequence, an altered helical period would not be unexpected, as pure poly(dA:dT) copolymers are known to have shorter helical periods (10.1 bp/turn) than random DNAs (10.6 bp/turn) \cite{Peck1981,Strauss1981}.  On the other hand, 5S exhibits the same helical period as E8 and TA in cyclization assays \cite{Cloutier2005}, so it is intriguing that its looping maximum occurs at a different length than that of E8 and TA, perhaps suggesting a different helical period in the context of looping than that of E8 and TA.  The promoter does not appear to alter the maximum of looping for a given sequence. As noted above, it is difficult to use these looping data to comment further on other DNA elasticity parameters, in particular any sequence-dependent differences in torsional stiffness, but in Fig.~\ref{fig:SITwistStiffness} in File S1 we provide evidence that these sequences may share the same twisting flexibility, even if they differ in helical period.

\begin{figure}[tbp]
  \begin{center}
   \includegraphics[width=6in]{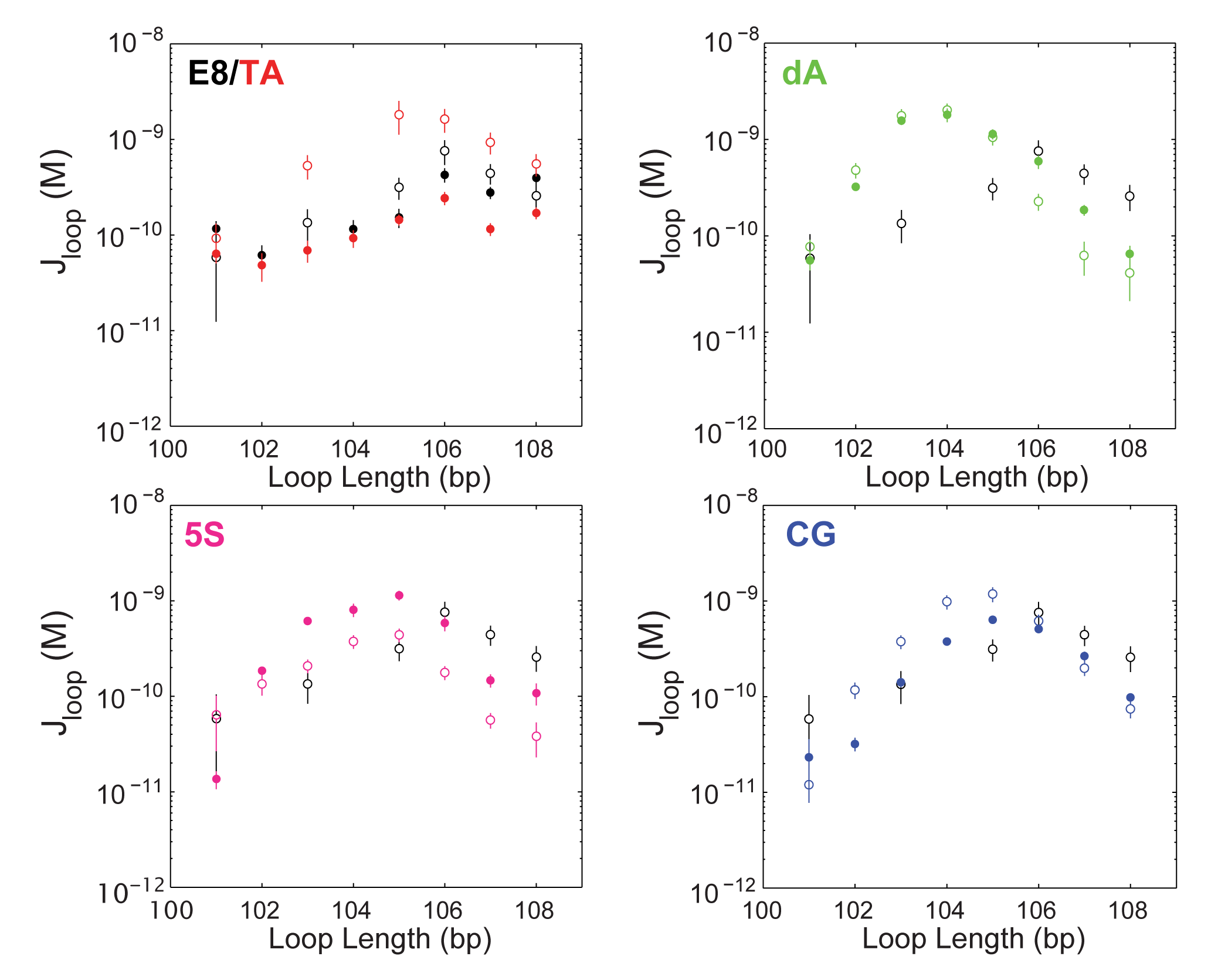}
   \end{center}
  \caption{
  {\bf Looping J-factors as a function of loop length and sequence.}  J-factors for sequences without (closed circles) and with (open circles) the \textit{lac}UV5 promoter were extracted from the data in Fig.~\ref{fig:PloopTot} as described in the Methods section.  The J-factor is a measure of the free energy of loop formation (and is related to the bending and twisting flexibility of the DNA in the loop): the higher the J-factor, the lower the free energy required to deform the loop region DNA into a loop.  As described in \cite{Johnson2012}, the addition of the promoter to the E8 loop sequence does not significantly affect its J-factor,  so the J-factor for E8 with the promoter is shown as a reference in all panels (black open points).  In contrast to E8, the addition of the promoter does change the J-factors for three of the four other sequences, making the TA-containing loops more flexible, but the 5S and, to a lesser extent,  CG sequences less flexible.  Interestingly, the poly(dA:dT) sequence, like the E8 sequence, is unchanged with the inclusion of the promoter.  We note that because the no-promoter versus with-promoter constructs contain different combinations of repressor binding sites, we can only use J-factors, not looping probabilities, to quantitatively examine the effect of the promoter; the statistical mechanical model of Eqn.~\ref{eqn:Ploop} allows us to make this comparison.}
  \label{fig:Jloop}
\end{figure}

The effect of the promoter on loop formation can be more clearly seen when looping J-factors are  compared across sequences, instead of the looping probabilities.  Because the no-promoter and with-promoter loops are flanked by different combinations of operators (Fig.~\ref{fig:PloopTot}(B); see also Methods), their looping probabilities cannot be directly compared.  However, as described above and in the Methods section, we can use the statistical mechanical model that we have described for this system to extract J-factors from each looping probability \cite{Johnson2012}.  These J-factors are shown in Fig.~\ref{fig:Jloop}.  Loop sequence can modulate the looping J-factor by at least an order of magnitude (compare the poly(dA:dT) J-factors to those of 5S with promoter or E8 and TA, no-promoter). The \textit{lac}UV5 promoter has the largest effect on the TA and 5S sequences (though of opposite sign), but appears to have little effect on poly(dA:dT)-containing and E8-containing loops, and moderate effect on CG-containing loops.  It is intriguing how large and diverse an effect the 36-bp \textit{lac}UV5 promoter has on the roughly 100 bp loops we examine here; but one possible explanation for its minimal effect on the poly(dA:dT)-rich sequence, at least, compared to the others, is that the properties of A-tract structures tend to dominate over the properties of surrounding sequences \cite{Haran2009}.  We note that our results in \cite{Johnson2012} comparing the effect of sequence versus flanking operators on  measured J-factors preclude the possibility that the differences between the no-promoter and with-promoter constructs are due to the difference in flanking operators.  We also note that it is possible that the effect of the promoter stems not from the promoter sequence itself, but from the fact that the sequences of interest that form the rest of the loop are shorter when 36 bp of the loop are replaced by the promoter sequence.  However, we consider this explanation to be less likely, because as shown in the left-hand panels of Fig.~\ref{fig:PloopTot}(C) and (D) above, we have measured the looping probabilities (and J-factors; see \cite{Johnson2012}) of more than two periods of E8- and TA-containing DNAs, allowing a direct comparison of loops that contain the same amount of E8 and TA both with and without the promoter (compare, for example, no-promoter loop lengths of 90 bp to with-promoter lengths of 120 bp).  In this case we still find that without the promoter the J-factors of the E8- and TA-containing loops are indistinguishable, but with the promoter the TA sequence loops more than the E8 sequence, indicating that it is the promoter and not a shortening of some unique element(s) of the E8 or TA sequences that cause the difference in J-factors with versus without the promoter for these two sequences.

\subsection{The Lac repressor supports a range of looped-state conformational preferences.}

TPM trajectories not only provide information about the free energy of loop formation, captured by the J-factors discussed in the previous section, but also contain some information about the preferred loop conformation as a function of sequence, through the observed length of the TPM tether when a loop has formed.  In fact, previous work from  our group and others has shown that the Lac repressor can support at least two observable loop conformations for any pair of operators, with any sequence, because these conformations lead to distinct tether lengths in TPM \cite{Johnson2012,Han2009,Edelman2003,Morgan2005,Wong2008,Normanno2008,Rutkauskas2009,Mehta1999,Morgan2005,Haeusler2012}. 
Although the underlying molecular details of these two looped states, which we label the ``middle'' (``M'') and ``bottom'' (``B'') states according to their tether lengths relative to the unlooped state, are as yet unknown, they must differ in repressor and/or DNA conformation in a way that alters the boundary conditions of the loop, since they are distinguishable in TPM.  It has been proposed that the two states arise from the four distinct DNA binding topologies allowed by a V-shaped Lac repressor similar to that shown in the Lac repressor crystal structure \cite{Towles2009,Lewis1996}, and/or two repressor conformations, the V-shape seen in the crystal structure and a more extended ``E'' shape \cite{Villa2005,Swigon2006,Zhang2006,Wong2008,Rutkauskas2009,Morgan2005,Haeusler2012}.  It is likely, in fact, that the two observed looped states are each composed of more than one microstate (that is, some combination of V-shaped and E-shaped repressor conformation(s) and associated binding topologies \cite{Towles2009}). Even without knowing the details of the underlying molecular conformation(s) of these two states, however, we can use them to provide a window into the effect of sequence on preferred loop conformation.

\begin{figure}[tbp]
  \begin{center}
    \includegraphics[width=6in]{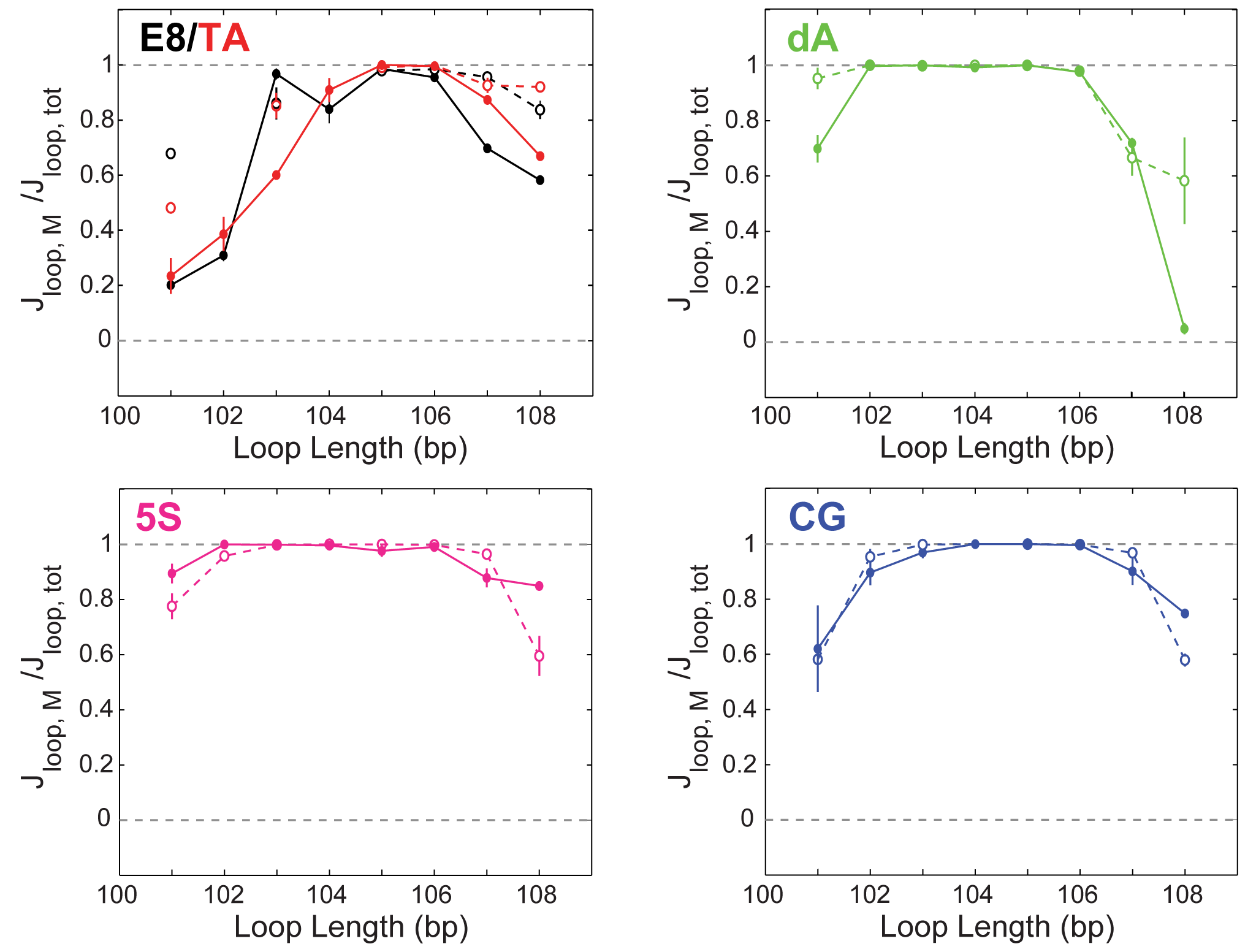}
    \end{center}
  \caption{{\bf Comparison of the likelihood of the ``middle'' (longer) versus ``bottom'' (shorter) looped states.  }The y-axes indicate the fraction of the total J-factor that is contributed by the middle state (as in Fig.~\ref{fig:Jloop}, since the with- and without-promoter constructs have different operators, J-factors and not looping probabilities must be compared).  That is, when the ratio $J_\mathrm{loop,M}/J_\mathrm{loop,tot}$ is unity, indicated by a horizontal black dashed line, only the middle state is observed; when this ratio is zero, again indicated by a horizontal black dashed line, only the bottom state is observed. Closed circles are no-promoter constructs; open circles are with-promoter.  E8 and TA data are a subset of those in \cite{Johnson2012}. Figure~\ref{fig:SIBvsM} in File S1 shows the looping probabilities and J-factors for the two states instead of the relative measures shown here.}
  \label{fig:PloopBvsM}
\end{figure}

In particular, by examining the relative probability of the two looped states as a function of both loop length and loop sequence, we can assess the contributions of sequence to the energy required to form the associated loop conformation(s).  As shown in Figure~\ref{fig:PloopBvsM}, which of the two looped states predominates depends in a complicated way upon the loop sequence, the presence versus absence of the  \textit{lac}UV5 promoter, and the loop length.  In \cite{Johnson2012}, we showed that having E8 or TA in the loop region, over two to three helical periods, leads to alternating preferences for the middle versus the bottom looped state, with the middle state predominating when the operators are in-phase and looping is maximal, but the bottom state predominating when the operators are out-of-phase.  The inclusion of the promoter in the loop increases the preference for middle state for out-of-phase operators.  These trends are captured in the top left panel of Fig.~\ref{fig:PloopBvsM}.  

These trends do not hold for the three genomically sourced loop sequences (CG, dA, and 5S).  For the poly(dA:dT)-rich sequence, as with E8 and TA, the promoter increases the preference for the middle looped state for out-of-phase operators; for 5S, however, the presence of the promoter decreases the preference for the middle state.  The preferred state of the CG sequence is mostly insensitive to the presence versus absence of the promoter.  Both with and without the promoter, though, the middle state is generally preferred ($J_\mathrm{loop, M}/J_\mathrm{loop, tot} \ge 0.5 $) at more loop lengths for the genomically sourced DNAs than for the synthetic sequences, insofar as we are able to determine from the lengths shown in Fig.~\ref{fig:PloopBvsM}.  
These results demonstrate a complicated dependence of preferred loop state on sequence that does not always follow overall trends in looping free energy: for example, 5S and TA are the two sequences that show the largest change in J-factor with the inclusion of the promoter, but E8 and TA are the sequences that show the largest change in preferred looped state with the promoter.  However, the trend seen in the preceding section with CG and poly(dA:dT) having more in common than 5S and TA holds true for preferred loop conformation as well.

\begin{figure}[tbp]
  \begin{center}
    \includegraphics[width=6in]{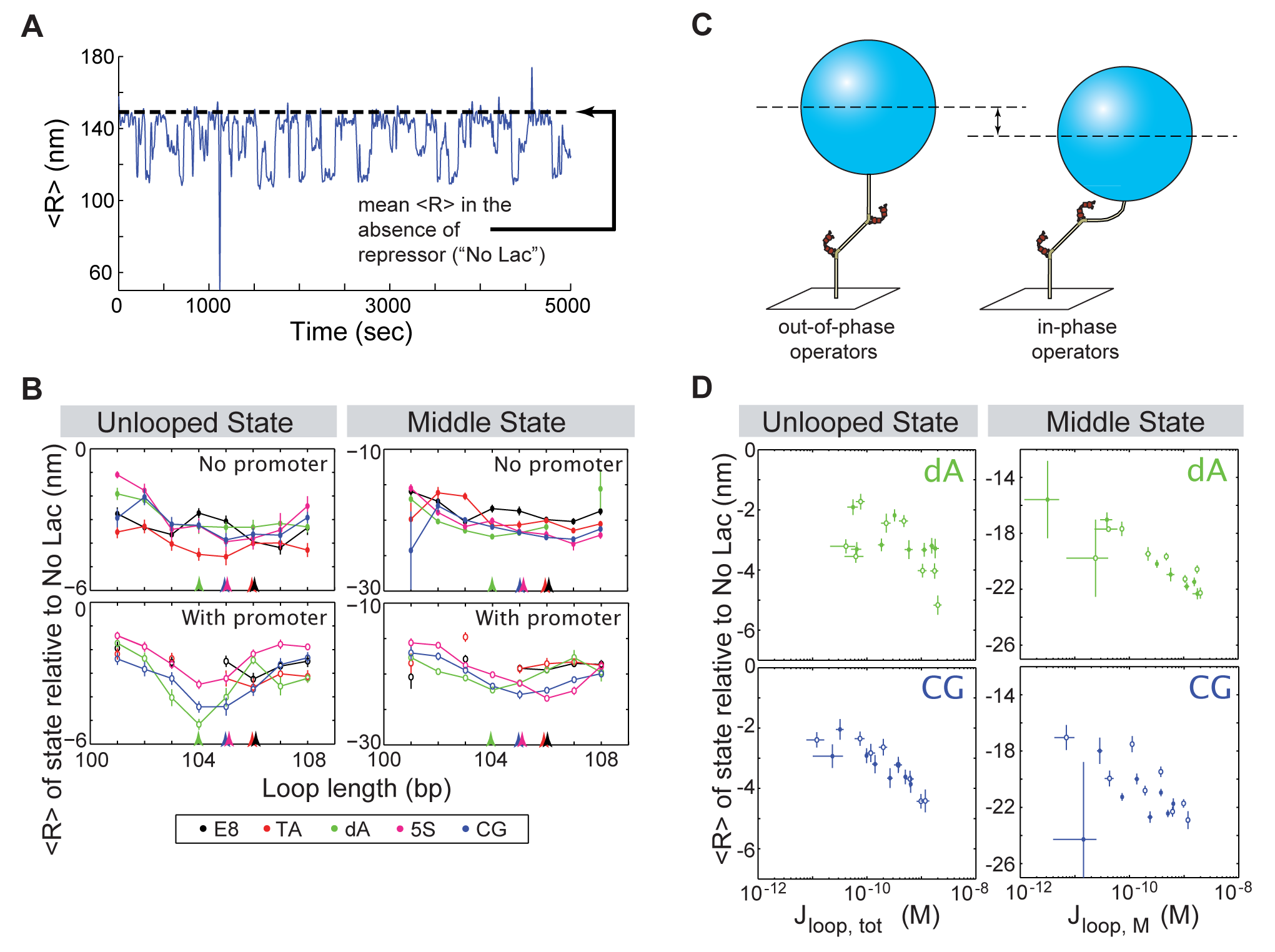}
    \end{center}
  \caption{{\bf Tether length as a function of loop length, sequence and J-factor.} {\bf (A)} Sample TPM time trajectory showing the smoothed (\textit{i.e.} Gaussian-filtered) root-mean-squared motion, $\langle R \rangle$, of a single bead.  This construct shows an unlooped state and two looped states, the ``middle'' state around 130 nm, and the ``bottom'' state around 110 nm.  Black horizontal dashed line indicates the average $\langle R \rangle$ for this particular tether in the absence of repressor. 
  Due to variability in tether length even in the absence of repressor \cite{Johnson2012}, on the y-axes in (B) and (D) we plot a relative measure of tether length, by normalizing the mean $\langle R \rangle$ value for a particular state to the mean $\langle R \rangle$ of each tether in the absence of repressor, and then taking the population average of this difference.   {\bf (B)}  Tether length as a function of loop length.  We observe a modulation of tether length with loop length, with the shortest tether lengths for both the unlooped and looped states occurring near the maximum of looping (indicated for each sequence by the colored arrows at the bottoms of the plots).   See Fig.~\ref{fig:SItetherlengths1} in File S1 for bottom state lengths. {\bf (C)} Schematic of our proposed model for the observed variations in unlooped tether length as a function of loop length, which we attribute to the phasing of the bends that the repressor creates upon binding the operators.  A similar argument can be made for the looped states.  Note that to emphasize the effect of bending from the operators, here we have for the most part represented the DNA as straight segments.
  {\bf (D)}  Tether length as a function of J-factor.  Unlooped state tether lengths are plotted versus the total J-factor, whereas middle state tether lengths are plotted versus the J-factor for the middle state.  As in (B), in general the length of the tether in both the unlooped and middle looped states is shorter at larger J-factors (that is, more in-phase operators) for a particular sequence.  However, this trend is sharper for some sequences  than others (see Fig.~\ref{fig:SItetherlengths2} in File S1 for the other sequences, which generally have more scatter than either the dA or CG sequences).}
  \label{fig:TetherLens}
\end{figure}

A different measure of loop conformation can be derived from the TPM tether lengths themselves---that is, from the measured root-mean-squared motion of the bead, $\langle R \rangle$, as in the example trajectory shown in Fig.~\ref{fig:TetherLens}(A), which exhibits three clear states, the two looped states and the unlooped state.   Because of variability in initial tether length, even in the absence of Lac repressor, we calculate a \textit{relative} measure of tether length for the unlooped and looped states, where the motion of each bead is normalized to its motion in the absence of repressor.  We might expect, then, that in the presence of repressor, the unlooped state would fall at a relative $\langle R \rangle$ of zero, and the looped states at negative values.  However, as can be seen in the sample trace in Fig.~\ref{fig:TetherLens}(A) and in the lefthand panels of Fig.~\ref{fig:TetherLens}(B), the unlooped state in the presence of repressor is actually shorter than the tether in the absence of repressor (\textit{i.e.,} the horizontal black dashed line in Fig.~\ref{fig:TetherLens}(A) lies above the mean of the unlooped state in the blue data).  In \cite{Johnson2012} we present evidence for this shortening of the unlooped state in the presence of repressor being due to the bending of the operators induced by the Lac repressor protein that is observed in the crystal structure of the Lac repressor complexed with DNA \cite{Lewis1996}.  (We note that this is a Lac repressor-specific result; compare, for example,  the recent results from Manzo and coworkers with the lambda repressor \cite{Manzo2012}, where a similar shortening of the unlooped state is shown to be due to nonspecific binding.  For example, the Lac repressor does not exhibit the dependence of the looped tether length on repressor concentration that is seen with the lambda repressor \cite{Johnson2012,Manzo2012}).  

As shown in Fig.~\ref{fig:TetherLens}(B), the length of the TPM tether in both the unlooped and looped states is similar but not identical for the five sequences and eight lengths that we examine here.  The most obvious modulation of tether length correlates with loop length, with the shortest unlooped- and looped-state tether lengths occurring near the maxima of the looping probability.  We believe this modulation with length is due to the phasing of the bends of the DNA tether as it exits the repressor-bound operators in the looped state, or the phasing of the bent operators in the unlooped state.
At the repressor concentration we use here, the unlooped state should be primarily composed of the doubly-bound state \cite{Johnson2012}, meaning that the two operators are both bent by bound repressor.  As shown schematically in Fig.~\ref{fig:TetherLens}(C), when these bends are in-phase, the tether length should be shortest (and also the looping probability is highest, because the operators are in-phase).  A similar argument can be made for the modulation of the looped state, regarding the relative phases of the tangents of the DNA exiting the loop.

  It is interesting to consider how the sequence of the loop might influence the length of the tether in the unlooped state, when no loop has formed; see, for example, the CG with-promoter versus 5S with-promoter sequences, where the latter is consistently longer than the former (Fig.~\ref{fig:TetherLens}(B)).  We do not see a sequence dependence to  tether length in the absence of repressor, ruling out the possibility of a detectable intrinsic curvature to the CG sequence.  We speculate instead that CG alters the trajectory of the DNA as it exits the bend in the operators in the unlooped state, compared to the trajectory when the sequence next to the operators is 5S, leading to a consistent difference in unlooped tether lengths.

Interestingly, in contrast to its influence on preferred looped state (middle versus bottom), the promoter does not alter the length of the tether for a given sequence at a given loop length (see also the bottom left panel of Fig.~\ref{fig:SItetherlengths1} and Fig.~\ref{fig:SItetherlengths2} in File S1).  On the other hand, as shown in Fig.~\ref{fig:TetherLens}(D), the poly(dA:dT)-rich sequence, noticeably more so than the other sequences, stands out as a sequence that does strongly affect the tether length of the loop, in that it mandates a very narrow range of tether lengths as a function of looping J-factor (related, for a particular sequence, to the loop length or equivalently the operator spacing). 
  A similar but less pronounced trend can be observed for the unlooped state with the GC-rich sequence  (Fig.~\ref{fig:TetherLens}(D)).   The other sequences allow much more variability in tether length as a function of J-factor/operator spacing (see Figure~\ref{fig:SItetherlengths2} in File S1).  This strong trend in tether length as a function of J-factor could be evidence of the formation of special, defined loop structures with the GC-rich and poly(dA:dT)-rich sequences that constrain the allowed loop conformations as a function of operator spacing more than the other sequences do.

Further computational and modeling efforts will be required to relate these data on tether lengths and preferred loop length to loop structure, similarly to how Towles and coworkers have used TPM tether lengths to show that different DNA loop topologies can explain the observed tethered lengths of the two looped states \cite{Towles2009}.  However, even without currently knowing the underlying molecular details causing these sequence-specific trends in tether length and preferred loop state, and therefore in loop conformation,   it is clear that it is the loop sequence, and not the Lac repressor itself, that determines the loop conformation to a large degree.  It has been shown recently that the Lac repressor is capable of accommodating many different loop conformations \cite{Haeusler2012}, which is consistent with the results we present here.  We hope that computational and modeling efforts with these data, as well as continued efforts to use assays such as FRET to directly probe loop conformation \cite{Mehta1999,Edelman2003,Morgan2005,Haeusler2012}, will shed light on this complex interplay between sequence and loop conformation.
 
\section{Discussion.}

\begin{figure}[!ht]
  \begin{center}
    \includegraphics[width=6in]{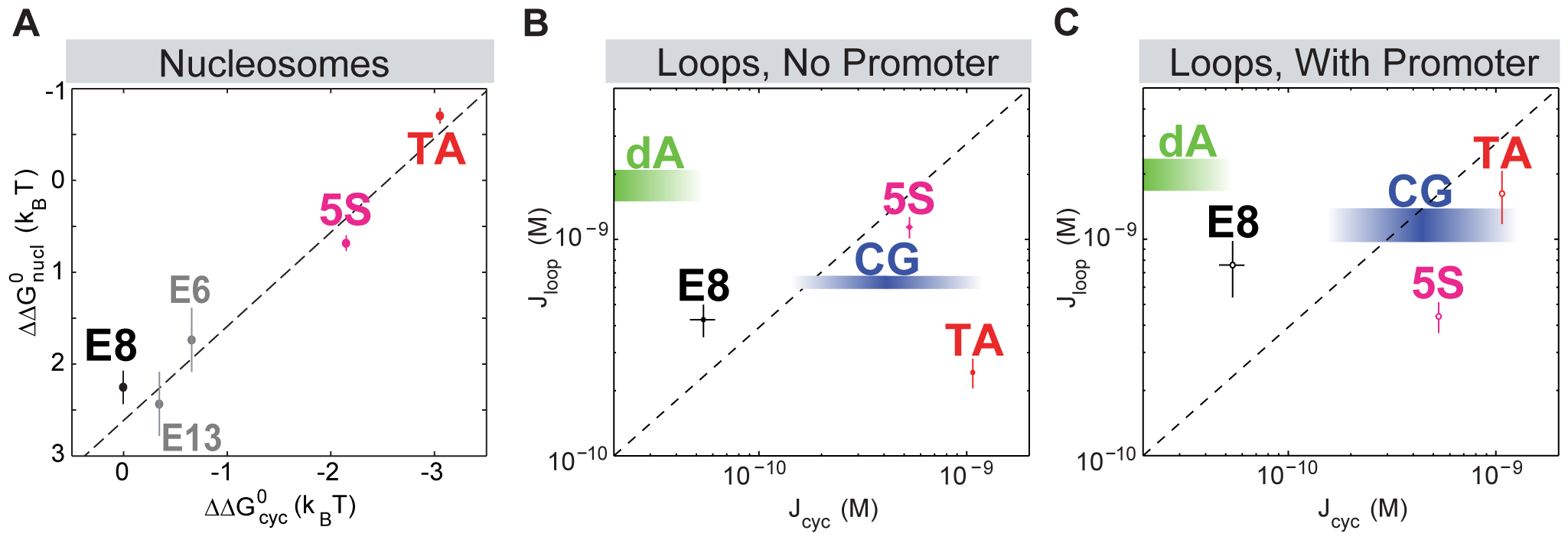}
   \end{center}
  \caption{{\bf Comparing trends in sequence flexibility for looping versus cyclization and nucleosome formation. } {\bf (A)} Nucleosome formation and cyclization share trends in sequence flexibility, with sequences that have lower energies of nucleosome formation ($\Delta\Delta G^0_{nucl}$) also having lower energies of cyclization ($\Delta\Delta G^0_{cyc}$).  Cloutier and Widom used this correlation to argue that the same mechanical properties, particularly the bendability, of the DNA contributed to nucleosome formation as to cyclization \cite{Cloutier2004}.  The energy of cyclization, $\Delta G^0_{cyc}$, is related to the cyclization J-factor for a particular DNA, $J_i$,  through the relationship $\Delta G^0_{cyc} = -RT \ln (J_i/J_{ref})$, where $T$ is the temperature, $R$ is the gas constant and $J_{ref}$ is an arbitrary reference molecule (see Ref.~\cite{Cloutier2004} for details).  Adapted from Refs.~\cite{Cloutier2004,Garcia2007}.  {\bf (B)} Looping J-factors for the no-promoter data do not show the same trends in sequence dependence as do cyclization and nucleosome formation: if anything, a higher cyclization J-factor correlates with a lower looping J-factor.  {\bf (C)} Same as (B) but for with-promoter DNAs.  The cyclization J-factors of the poly(dA:dT)-rich and GC-rich sequences that we use here have not been reported, so they are shown as shaded regions whose height reflects the uncertainty in the looping J-factors we measure, and whose width reflect our estimates about what their cyclization behavior should be.  In particular, the poly(dA:dT)-rich sequence exhibits very low nucleosome occupancy \textit{in vivo} \cite{Yuan2005,Field2008}, and similar sequences have high energies of nucleosome formation \textit{in vitro} \cite{Anderson2001,Field2008}, which, according to the logic of (A), should correspond to a low cyclization J-factor, probably lower than that of E8.  Some poly(dA:dT)-rich DNAs were in fact recently directly shown to cyclize less readily than random sequences \cite{Vafabakhsh2012}.  In contrast, the GC-rich sequence should be a good nucleosome former (though the nucleosome affinity of this particular sequence has not been tested either \textit{in vivo} or \textit{in vitro}), and so its cyclization J-factor is probably comparable to that of 5S and TA, the other strong nucleosome-preferring sequences on this plot.  Additional details of how this plot was generated can be found in the Methods section.}
  \label{fig:CompareCycLoop}
\end{figure}

In \cite{Johnson2012} we showed that the synthetic E8 and TA sequences show no sequence dependence to looping in the absence of the \textit{lac}UV5 promoter but a nucleosome-like sequence dependence in the presence of the promoter.  We hypothesized that perhaps the promoter alters the preferred state of the loop to one whose shape is more similar to that of DNA in a nucleosome or DNA minicircle formed by cyclization, leading to similar sequence  trends with the promoter as with nucleosomes.  We still attribute the difference in the patterns of sequence dependence that we observe between looping and nucleosome formation to the role of the shape of the deformation in determining the observed deformability of a particular sequence. 
However, we have  shown here with a broader range of sequences that the role of the promoter in controlling loopability is more complicated than we had previously hypothesized.  Neither with nor without the promoter does loop formation follow the sequence trends of nucleosome formation.  As shown in Figure~\ref{fig:CompareCycLoop}, if looping J-factors did follow the same patterns of sequence preference as do cyclization J-factors and nucleosome formation free energies, a plot of the looping J-factors versus cyclization J-factors for the various sequences we have studied here would fall on a line with a positive slope.  We find that this is not the case; in fact, without the promoter there is perhaps a slight anticorrelation between looping J-factors and cyclization J-factors (and no discernible correlation with the promoter). 

The strong correlation between a sequence's ease of cyclization and of nucleosome formation, as shown in Fig.~\ref{fig:CompareCycLoop}(A), has been used to argue that nucleosome sequence preferences depend largely on the intrinsic mechanical properties of a DNA, particularly its bendability \cite{Cloutier2004}, though other mechanisms have also been proposed, such as that described by Rohs and coworkers, which depends not on sequence-dependent DNA flexibility but on sequence-dependent minor groove shape \cite{Rohs2009}.  We have shown here that three sequence features that commonly determine nucleosome preferences, either through their effect on DNA flexibility or on other structural aspects recognized by the nucleosome, do not likewise determine looping, arguing for the need to identify a different set of sequence features that determine loopability. The most striking contrast between previously established sequence ``rules'' derived from nucleosome studies and the trends in looping J-factors that we observe here is that of the nucleosome-repelling, poly(dA:dT) sequence, which has the lowest looping free energy that we have quantified so far.  Other \textit{in vitro} assays predominantly show poly(dA:dT) copolymers to be highly resistant to deformations; for example, Vafabakhsh and coworkers recently used a FRET-based cyclization assay, analogous to traditional ligase-mediated cyclization assays, to show that poly(dA:dT)-rich sequences have cyclization rates significantly smaller than other sequences such as E8 and TA \cite{Vafabakhsh2012}.  Although ease of cyclization is often equated with bendability, it appears that such observed bendability is more context-dependent than has been previously appreciated: that is, the simplest model that one would write down to describe the energetics of these different deformed DNAs would  feature the persistence length as the governing parameter that is used to characterize bendability, and yet, the distinct responses seen in looping, nucleosomes and cyclization belie that simplest model.   It will be informative to extend this study of an unphased poly(dA:dT) tract in DNA loops to include more sequences containing both pure poly(dA:dT) copolymers and naturally-occuring poly(dA:dT)-rich DNAs that exclude nucleosomes \textit{in vivo},  in order to elucidate the precise role of poly(dA:dT)-tracts in determining looping.  It is clear, however, that poly(dA:dT)-rich DNAs should not be exclusively thought of as stiff or resistant to bending in all biological contexts.

\begin{figure}[tbp]
  \begin{center}
    \includegraphics[width=3in]{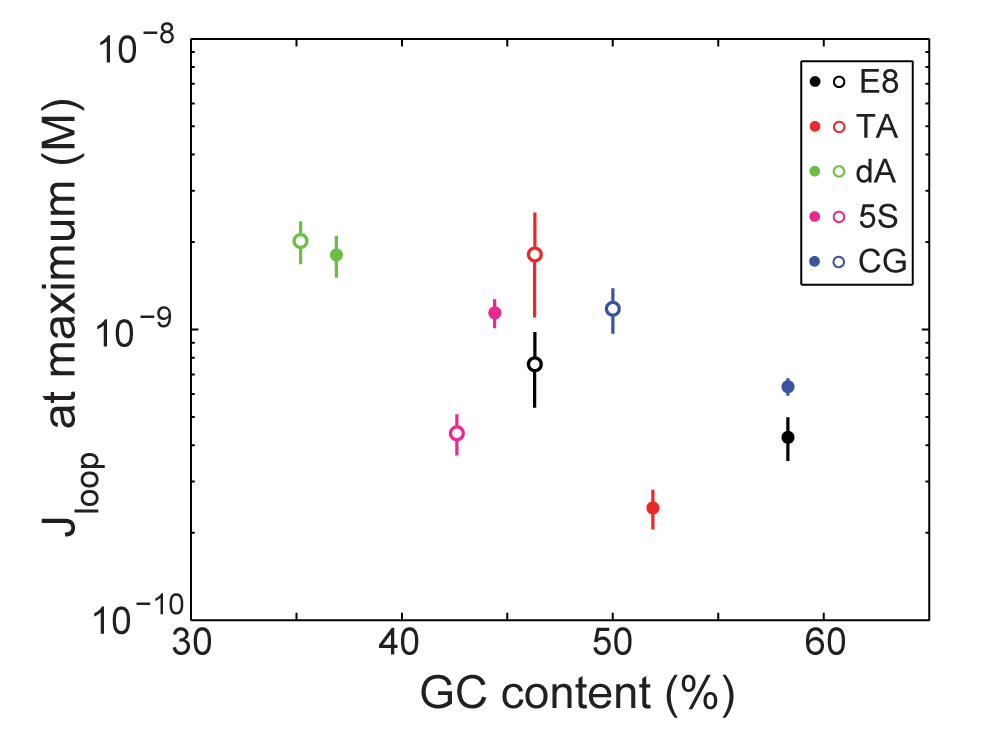}
    \end{center}
  \caption{{\bf Maximum looping J-factor as a function of loop G+C content.}  Maximum J-factors for each of the five sequences, with (closed circles) and without (open circles) promoter, are plotted with respect to each sequence's G+C content.  
  For nucleosomes, G+C content strongly correlates with nucleosome occupancy \cite{Tillo2009}.  In contrast,  it appears that G+C content and loopability are anticorrelated. 
  Loop lengths plotted here are the same as in Fig.~\ref{fig:CompareCycLoop}.}
  \label{fig:GCcontent}
\end{figure}
 
A second striking contrast between our results here and previously established rules for nucleosome formation concerns the role of G+C content in determining loop formation.  The G+C content of a DNA is one of the most powerful parameters for predicting nucleosome occupancy \textit{in vivo} \cite{Tillo2009,Tillo2010}, with higher G+C content correlating with higher occupancy.  
However, as shown in Fig.~\ref{fig:GCcontent}, G+C content offers little predictive power for loopability, or is anticorrelated with looping.  We note that a recent, systematic DNA cyclization study demonstrated a quadratic dependence of DNA bending stiffness on G+C content \cite{Geggier2010}. In our case of protein-mediated DNA looping, the looping J-factor contains contributions from protein elasticity in addition to those from DNA elasticity, and our DNA sequences contain A-tracts and GGGCCC motifs that were excluded in \cite{Geggier2010}, making a direct comparison between our results and theirs difficult; but it is possible that the looping J-factor is neither correlated or anticorrelated with G+C content but instead depends quadratically on G+C content, as do cyclization J-factors. More data will be necessary to make a strong statistical statement about the anticorrelation or lack of correlation between the looping J-factor and G+C content, and to determine the form of the relationship between the looping J-factor and G+C content (\textit{e.g.} quadratic versus linear), but we propose low G+C content as the starting point of a potential new sequence ``rule'' for predicting looping J-factors, and a fertile realm of further investigation.  Finally, we have shown that the repeating AA/TT/TA/AT and GG/CC/GC/CG steps that characterize the 5S and TA sequences, as well as many nucleosome-preferring sequences, do not likewise determine looping J-factors, as these two sequences behave very differently from each other in the context of transcription factor-mediated DNA looping.

\section{Conclusions}

Here we have extended our previous work on the sequence dependence of loop formation by the Lac repressor to include three naturally occurring, genomic sequences that have either nucleosome-repelling or nucleosome-attracting functions \textit{in vivo}, in addition to the two synthetic sequences we described previously \cite{Johnson2012}.  We find that two sequences that share sequence features important to nucleosome formation and that share trends in observed flexibility in cyclization and nucleosome formation assays, the 601TA and 5S sequences, behave less similarly in the context of DNA looping than the two sequences that should have least in common, the GC-rich, nucleosome attracting sequence and the poly(dA:dT)-rich, nucleosome repelling sequence.  5S and TA share neither trends in looping free energy relative to the random E8 sequence, nor loop length where looping is maximal, nor preferred loop conformation, nor their response to the larger sequence context (as evidenced by the fact that the inclusion of the \textit{lac}UV5 promoter sequence in the loop increases the looping J-factor for TA but decreases it for 5S).  

We have also shown that a poly(dA:dT)-rich DNA that forms a nucleosome-free region in yeast \cite{Yuan2005} is actually extremely deformable in the context of looping by a transcription factor.  The rest of the sequences show a range of J-factors that does not correlate with any observed trends in flexibility as measured by ligase-mediated cyclization assays, nor with the observation that high G+C content correlates with nucleosome occupancy \cite{Tillo2009}.  The diversity of the effects on DNA looping that we observe with these five sequences (ten, if the inclusion of the promoter is considered to create a ``new'' sequence) underscores the necessity of a large-scale screen for sequences that control loop formation both \textit{in vivo} and \textit{in vitro}, much as has been done in the context of nucleosome formation to help establish the sequence-dependence rules of that field (for example, see \cite{Lowary1998,Widlund1999}).

Our work in no way undermines previous claims of the sequence dependence to nucleosome formation and/or occupancy either \textit{in vivo} or \textit{in vitro}; rather, it demonstrates that the ``rules'' of sequence flexibility derived from cyclization and nucleosome formation studies are inapplicable to DNA looping, possibly due to the difference in the boundary conditions and therefore DNA conformations involved in forming a protein-mediated loop versus a DNA minicircle or a nucleosome.  
It will  be interesting to extend these studies of the role of sequence in loop formation to other DNA looping proteins besides the Lac repressor. As noted above, it has been shown recently that the Lac repressor can accommodate many different loop conformations \cite{Haeusler2012}.  The variety in tether lengths and preferred looped states that we observe are consistent with a forgiving Lac repressor protein. Nucleosomes, on the other hand, have a more fixed structure that should not be as accommodating to a range of helical periods and DNA polymer conformations (hence the hypothesis that poly(dA:dT)-rich DNAs disfavor nucleosome formation because they adopt geometry that is incompatible with the structure of the DNA in a nucleosome \cite{Segal2009}). It would be informative to measure the looping J-factors of these same sequences with a more rigid looping protein. It will also be interesting to see if other bacterial promoter sequences have similar effect of altering the looping boundary condition as the very strong and synthetic \textit{lac}UV5 promoter.  In fact, the \textit{lac}UV5 promoter should be a key starting point for identifying sequences that have a strong effect on looping, since it can have significant effects on the behavior of a loop, even when it comprises only one-third of the loop length.

\section{Materials and Methods}

\subsection{DNAs.}

The poly(dA:dT)-rich sequence (from Fig. 4 of Ref. \cite{Yuan2005}), GC-rich sequence (from ``Human 2'' at http://genie.weizmann.ac.il/pubs/field08/field08\_data.html), and 5S sequences (from Fig. 1 of \cite{Simpson1983}) were cloned into the pZS25 plasmid used in \cite{Johnson2012}, with these eukaryotic sequences replacing the E8 or TA sequences in that plasmid.  In cases where the loop lengths used in this study were shorter than the 147  bp that are wrapped in nucleosomes, the corresponding looping sequences used in TPM were taken from the middle of these sequences (relative to the nucleosomal dyad); in cases where the nucleosomal sequences were shorter than the desired loop length, they were padded at one end with the random E8 sequence \cite{Cloutier2004,Cloutier2005,Johnson2012}.  See Figures~\ref{fig:SIseqlist1} and~\ref{fig:SIseqlist2} in File S1 for details.  As in \cite{Johnson2012}, ``no-promoter'' loops were flanked by the synthetic, strongest known operator (repressor binding site) $O_{id}$ and the strongest naturally occurring operator $O_1$; ``with-promoter'' loops were flanked by $O_{id}$ and a weaker naturally occurring operator, $O_2$, because these with-promoter constructs are also used in \textit{in vivo} studies of the effect of loop architecture on YFP expression, in which case $O_2$ is a more convenient choice of operator than $O_1$.  Similarly, the motivation to include the \textit{lac}UV5 promoter in the loop stems from parallel \textit{in vivo} studies, in which the promoter is a natural part of the looping architecture.  The promoter is included in the loop between the sequence of interest and the $O_2$ operator.  Figures~\ref{fig:SIseqlist1} and~\ref{fig:SIseqlist2} in File S1 gives the exact sequences used in this work; Fig.~\ref{fig:PloopTot}(B) shows the TPM constructs schematically.

Cloning of the sequences of interest into the pZS25 plasmid was accomplished in either one or two steps.  For the 5S sequences, oligomers were first ordered from Integrated DNA Technologies as single-stranded forward and reverse complements, consisting of 69 bp (for the ``with-promoter'' constructs) or 105 bp (for the ``no-promoter'' constructs) of the 5S sequence, plus the $O_{id}$ and $O_1$/$O_2$ operators, and, where applicable, the \textit{lac}UV5 promoter sequence.  These oligomers were annealed and then ligated into the pZS25 plasmid at the AatII and EcoRI restriction sites that fall just outside the operators that flank the E8 or TA sequences in the original pZS25 plasmids \cite{Johnson2012}.  Second, Quik-Change mutagenesis (Agilent Technologies) was performed to generate additional lengths (that is, to introduce insertions or deletions) of the 5S sequence from the initial 105 bp loop lengths.  However, we found that this site-directed mutagenesis step generated distributions of products for the poly(dA:dT) constructs, possibly due to replication slipped mispairing over repetitive sequences \cite{Bzymek2001}.  Therefore all lengths of the poly(dA:dT) sequence, as well as of the GC-rich sequence, which also have the potential to contain such ``slippery'' regions, were created by ligation of synthesized oligomers into the pZS25 plasmid.   All constructs were confirmed by sequencing (Laragen Inc.) to have clean sequence reads, and the approximately 450~bp digoxigenin- and biotin-labeled TPM constructs were created by PCR as described for the E8- and TA-containing constructs in \cite{Han2009,Johnson2012}.  Sequences of TPM constructs were again confirmed by sequencing before use. 

\subsection{TPM sample preparation, data acquisition and analysis.}

Tethered particle motion assays were performed as described in \cite{Johnson2012}.  Briefly, linear DNAs, labeled on one end with digoxigenin and on the other end with biotin, were introduced into chambers created between a microscope slide and coverslip, with the coverslip coated nonspecifically with anti-digoxigenin.  Streptavidin-coated beads (Bangs Laboratories, Inc) were then introduced into the chamber to complete the formation of tethered particles.  The motion of the beads was tracked using custom Matlab code that calculated each bead's root-mean-squared (RMS) motion in the plane of the coverslip, and looping probabilities were extracted from these  RMS-versus-time trajectories as the time spent in the looped state (reduced RMS), divided by total observation time.  Similarly, the probabilities of the ``bottom'' versus ``middle'' states (see Results section) were defined as the time spent in a particular state, divided by the total observation time.

By measuring the looping probability of a construct at a particular repressor concentration, and using the repressor-operator dissociation constants for $O_1$, $O_2$ and $O_{id}$ in \cite{Johnson2012}, we can calculate the J-factor for that construct.  All measurements in this work were carried out at 100 pM repressor, using repressor purified in-house. The relationship between the looping probabilities measured in TPM ($p_\mathrm{loop}$), the repressor-operator dissociation constants for the two operators that flank the loop ($K_1$, $K_2$ and $K_{id}$), and the looping J-factor of the DNA in the loop ($J_\mathrm{loop}$) can be described as
\begin{equation}
p_{\mathrm{loop}}
=\frac{\frac{[R]J_{\mathrm{loop}}}{2K_{A}K_{B}}}{1+\frac{[R]}{K_{A}} + \frac{[R]}{K_{B}}+ \frac{[R]^2}{K_{A}K_{B}} +\frac{[R]J_{\mathrm{loop}}}{2K_{A}K_{B}}},\label{eqn:Ploop}
\end{equation}
where $[R]$ is the concentration of Lac repressor, and $K_{A}$ and $K_{B}$ are repressor-operator dissociation constants of the two operators flanking the loop ($K_{id}$ and $K_1$ or $K_2$). 
A similar expression can be derived for the J-factors of the individual ``bottom'' and ``middle'' looped states and is given in \cite{Johnson2012}.

\subsection{Generating the plots in Figure~\ref{fig:CompareCycLoop}.}

The J-factors plotted in Figure~\ref{fig:CompareCycLoop} are the maximum looping or cyclization J-factors over a particular period.  Specifically, the looping J-factors used are those at 104 bp for dA, 105 for 5S and CG, and 106 for E8 and TA; the cyclization J-factors are for 94 bp of the E8, 5S or TA sequences and are taken from \cite{Cloutier2005}.  Although we are not directly comparing identical lengths between cyclization and looping, the general trends hold regardless of lengths chosen.  In fact, identifying the loop length that corresponds to a particular cyclization length is difficult, given that the flanking operators for looping must be taken into account in some fashion.  That is, for cyclization, DNA length is easy to compute---it is simply the length of the oligomer used in the ligation reactions.  However, in the case of looping, it is unclear if the appropriate length for comparison is just the DNA in the loop (excluding the operators), or the length between the midpoints of the operators, or including all of the operators. Similarly, we are not comparing identical loop lengths across sequences; we chose to compare loop flexibility at the looping maximum for each sequence in an attempt to compare lengths at which the operators are most likely to be in phase, such that we are comparing only bending and not twisting flexibility. Finally, we note that here we are interested in the same kind of comparison that Cloutier and Widom were in Ref.~\cite{Cloutier2004}, which was the inspiration for this figure; in~\cite{Cloutier2004}, Cloutier and Widom compared cyclization and nucleosome formation free energies, even though the cyclization experiments were performed with roughly 100 bp DNAs and the nucleosome formation assays with roughly 150 bp DNAs.  Likewise, we do not  expect that the fragments of nucleosome-preferring or nucleosome-repelling sequences that we examine here in the context of looping will necessarily have exactly the same characteristics as the full-length nucleosomal sequences from which they were derived; but we are interested in comparing general trends in observed flexibility of these roughly 110 bp loops with those of roughly 100 bp ligated minicircles and  of roughly 150 bp nucleosomal DNAs.

\section{Acknowledgements}

We are indebted to the late Jon Widom for the inspiration of this project and for his guidance, mentorship and friendship over many years.  We thank Chao Liu, David Wu, David Van Valen, Hernan Garcia, Martin Lind\'{e}n, Mattias Rydenfelt, Yun Mou, Tsui-Fen Chou, Eugene Lee, Matthew Raab, Daniel Grilley, Niv Antonovsky, Lior Zelcbuch, Matthew Moore, Ron Milo, Eran Segal, and the Phillips, Mayo, Pierce and Elowitz labs for insightful discussions, equipment and technical help; and Winston Warman at Transgenomic, Inc. (Omaha, NE, USA) and Jin Li at Laragen, Inc (Culver City, CA, USA) for special help with sequencing the poly(dA:dT)-rich DNAs. 

\section{Supporting Information}

{\bf File S1: Supporting figures.}  Figure S1 ``No-promoter'' looping sequences used in this work.  Figure S2 ``With-promoter'' looping sequences used in this work.  Figure S3 Sequence-dependent twist stiffness.  Figure S4 Looping probabilities and J-factors for the two looped states separately.  Figure S5 Tether lengths of looped and unlooped states as a function of loop length and sequence.  Figure S6 Tether length as a function of J-factor.


\end{document}